\documentclass[12pt]{article}
\usepackage{amsmath}
\usepackage{cite}

\csname @addtoreset\endcsname{equation}{section}
\newcommand{\eq}{\begin{equation}}
\newcommand{\feq}{\end{equation}}
\newcommand{\eqn}{\begin{eqnarray}}
\newcommand{\feqn}{\end{eqnarray}}
\newcommand{\arr}{\begin{eqnarray*}}
\newcommand{\farr}{\end{eqnarray*}}

\newcommand{\fft}[2]{{\frac{#1}{#2}}}
\newcommand{\ft}[2]{{\textstyle\frac{#1}{#2}}}
\long\def\omit#1{}

\newcommand\HH{\mathcal{H}}

\newcommand\MM{\mathcal{M}}
\newcommand\RR{\mathcal{R}}
\newcommand\OO{\mathcal{O}}

\newcommand\GG{\mathcal{G}}

\newcommand\dM{\partial \MM}

\newcommand\nts{\negthickspace}

\newcommand\tg{h}

\newcommand\vf{\phi}

\DeclareMathAlphabet{\mathpzc}{OT1}{pzc}{m}{it}

\begin{document}
\begin{titlepage}
\begin{flushright}
CAMS/04-03\\
MCTP-04-49\\
hep-th/0408205
\end{flushright}

\vspace{.3cm}
\begin{center}

\renewcommand{\thefootnote}{\fnsymbol{footnote}}
{\Large \bf Black hole mass and Hamilton-Jacobi counterterms}\\

\vspace{1cm}

{\large \bf {A. Batrachenko$^1$, James T. Liu$^1$,
R. McNees$^1$, W.~A.~Sabra$^2$\\[.2cm]
and W. Y. Wen$^1$\footnote{Emails:
\vtop{\hbox{abat@umich.edu, jimliu@umich.edu, ramcnees@umich.edu,
ws00@aub.edu.lb,}\hbox{wenw@umich.edu}}}}}\\[.3cm]

\renewcommand{\thefootnote}{\arabic{footnote}}
\setcounter{footnote}{0}
{\small
$^1$ {Michigan Center for Theoretical Physics,\\
Randall Laboratory of Physics, The University of Michigan,\\
Ann Arbor, MI 48109--1120, USA}\\
\vspace*{0.4cm}
$^2$ Center for Advanced Mathematical Sciences (CAMS)\\
and\\
Physics Department, American University of Beirut, Lebanon.}
\end{center}

\vspace{1.5cm}
\begin{center}
{\bf Abstract}
\end{center}
We apply the method of holographic renormalization to computing black
hole masses in asymptotically anti-de Sitter spaces.  In particular, we
demonstrate
that the Hamilton-Jacobi approach to obtaining the boundary action yields
a set of counterterms sufficient to render the masses finite for four,
five, six and seven-dimensional $R$-charged black holes in gauged
supergravities.  In addition, we prove that the familiar black hole
thermodynamical expressions and in particular the first law continues
to holds in general in the presence of arbitrary matter couplings to
gravity.

\end{titlepage}

\section{Introduction}

The AdS/CFT conjecture, where gravity in anti-de Sitter space is
holographically dual to a conformal field theory on the boundary,
has led to additional interest in black hole thermodynamics.  In this
context the thermal properties of an AdS black hole configuration are
dual to that of the finite temperature CFT.  A particularly well
studied example of this is the Hawking-Page phase transition
\cite{Hawking:1982dh} for black holes in AdS, corresponding to a
deconfinement transition in the dual field theory \cite{Witten:1998zw}.

A common approach to extracting thermodynamic quantities from the
black hole background is to evaluate the on-shell gravitational
action, $I$, as well as the boundary stress tensor $T^{ab}$, given by
\begin{equation}
T^{ab}=\fft2{\sqrt{-h}}\fft{\delta I}{\delta h_{ab}},
\end{equation}
where $h_{ab}$ is the boundary metric.  According to black hole
thermodynamics, the on-shell value of the action may be identified
with the thermodynamic potential $\Omega$ according to $I=\beta
\Omega$.  For static backgrounds with the time-like Killing
vector $\partial/\partial t$, the energy $E$ is given by the ADM
mass, extracted from the $tt$ component of the boundary stress tensor.

Black hole thermodynamics has been widely explored in the context
of pure Einstein gravity with a cosmological constant.
In this case it is well known that the first law of thermodynamics,
$dE=T\,dS$, holds rather generally.  Furthermore, the thermodynamic
potential $\Omega$ is equivalent to the Helmholtz free energy $F$,
so that $F=E-TS$ is also satisfied.  The fact that these
features of black hole thermodynamics closely parallel those of ordinary
thermodynamics has been the motivation behind the study of AdS/CFT at
finite temperature.

For AdS/CFT, however, it is necessary to extend the results
of black hole thermodynamics to encompass gauged supergravities or
more general systems of matter coupled to gravity.
In these systems it is possible to turn
on conserved $R$-charges in addition to the temperature. 
It is then appropriate to work with the grand
canonical ensemble, as discussed in~\cite{Cvetic:1999ne}. 
The thermodynamic potential is related to the energy according to
\begin{equation}
\Omega=E-TS-\Phi^IQ_I,
\label{eq:thermoi}
\end{equation}
where $Q_I$ are the set of conserved $R$-charges and $\Phi^I$ are the
corresponding electric potentials (which play the role of chemical
potentials).

Although it is expected that (\ref{eq:thermoi}) would be satisfied
in general, a slight complication arises in that both $\Omega$ and
$E$ (extracted from the on-shell Euclidean action and the boundary
stress-tensor, respectively) are divergent quantities and require
renormalization.  One approach to dealing with this problem, as suggested
by Brown and York \cite{Brown:1992br}, is to subtract the divergent action
of a reference spacetime from the action for the spacetime of interest. In
many cases this technique is sufficient, but it suffers from two main
drawbacks. First, it requires that we embed a boundary with intrinsic
metric $h_{ab}$ in the reference spacetime, which is often not possible.
Second, the procedure is not intrinsic to the spacetime of interest, and
all physical quantities are defined with respect to a particular reference
spacetime. This becomes problematic when the appropriate reference
background is unknown or ambiguous.

These problems can be avoided by using the boundary counterterm approach
for removing divergences from the action
\cite{Balasubramanian:1999re,Emparan:1999pm}.  There are two common
prescriptions for calculating the boundary counterterms.  The first of
these involves the asymptotic expansion of bulk fields near the boundary
of spacetime. This approach is clearly defined and rigorous; it provides
a complete set of covariant counterterms that remove all divergences
from the on-shell action
\cite{Henningson:1998gx,Kraus:1999di,%
Taylor-Robinson:2000xw,deHaro:2000xn,Skenderis:2000in,Bianchi:2001de,%
Bianchi:2001kw,Skenderis:2002wp}. 
The second method, which we employ in this paper, is based on the
Hamilton-Jacobi formalism
\cite{deBoer:1999xf,deBoer:2000cz,Kalkkinen:2001vg,Martelli:2002sp,%
Larsen:2003pf, Larsen:2004kf}.
The Hamilton-Jacobi formalism, which has found many applications in
semi-classical gravity, was first applied in the AdS/CFT context by de
Boer, Verlinde, and Verlinde \cite{deBoer:1999xf}. We will not discuss the
motivations and subtleties of this approach; instead we refer the reader
to the excellent review by M\"uck and Martelli in \cite{Martelli:2002sp}.

In this paper we reexamine the familiar asymptotically AdS$_d$ black
hole solutions of gauged supergravities in 4, 5, 6 and 7 dimensions, and
demonstrate how divergences are renormalized through the addition of
appropriate Hamilton-Jacobi counterterms.  Given a well-defined
renormalization scheme, we are able to prove that the relation
(\ref{eq:thermoi}) is automatically satisfied for all such black
hole solutions.  This proof is, in fact, quite general, and is
anticipated to remain valid for more general asymptotically AdS$_{d}$
backgrounds.

\section{General Charged Black Hole Solutions}

While details of the various supergravity theories depend crucially on
dimension, general features of the bosonic sector can be treated in a
dimension independent manner.  We may thus consider a general bosonic
action for gravity coupled to a set of scalar and vector fields given
in the form
\begin{eqnarray}
I[g_{\mu\nu},\phi^i,A_\mu^I]&\!\!=\!\!&-\fft1{16\pi G_d}
\int_{\mathcal{M}}d^dx\sqrt{-g}\;[R-\ft12{\cal G}_{ij}(\phi)
\partial_\mu\phi^i\partial^\mu\phi^j-\ft14G_{IJ}(\phi)F_{\mu\nu}^I
F^{\mu\nu\,J}-V(\phi )] \notag \\
&&+\fft1{8\pi G_d}\int_{\partial\mathcal{M}}d^{d-1}x\sqrt{-h}\;\Theta .
\label{eq:bulka}
\end{eqnarray}
This action is appropriate to a $d$-dimensional spacetime ${\cal M}$ with
a $(d-1)$-dimensional boundary $\partial{\cal M}$. The Gibbons-Hawking
surface term is given in terms of the trace of the extrinsic curvature
$\Theta_{\mu\nu}$ of the boundary
\begin{equation} \label{ext_curv_defn}
\Theta_{\mu\nu}=-\ft12(\nabla_\mu n_\nu+\nabla_\nu n_\mu),
\end{equation}
where $n^\mu$ is the outward-pointing normal on $\partial{\cal M}$, and
$h_{\mu\nu}$ is the induced metric.  The equations of motion derived from
(\ref{eq:bulka}) are
\begin{eqnarray}
&&R_{\mu \nu }={\frac{1}{2}}{\cal G}_{ij}(\phi)
\partial _{\mu }\phi ^{i}\partial
_{\nu }\phi ^{j}+{\frac{1}{2}}G_{IJ}
\left(F_{\mu \lambda }^{I}F_{\nu }{}^{\lambda\,J}
-\fft1{2(d-2)}g_{\mu\nu}F_{\rho\sigma}^IF^{\rho\sigma\,J}\right)
+\fft1{d-2}g_{\mu \nu }V,\nonumber\\
&&\nabla^\mu(G_{IJ}F_{\mu\nu}^J)=0,\nonumber\\
&&\nabla^\mu({\cal G}_{ij}\nabla_\mu\phi^j)=\fft12
(\partial_{\phi^i}{\cal G}_{jk})\partial_\mu\phi^j\partial^\mu\phi^k
+\fft14(\partial_{\phi^i}G_{IJ})F_{\mu\nu}^IF^{\mu\nu\,J}+\partial_{\phi^i}V.
\label{eq:eins}
\end{eqnarray}

Since we are interested in spherically symmetric black holes carrying
electric charge, in much of the following we choose to work with a field
ansatz of the form
\begin{eqnarray}
&&ds^{2}=-e^{-2(d-3)B(r)}f(r)dt^{2}+e^{2B(r)}\left({\frac{dr^{2}}{f(r)}}
+r^{2}d\Omega _{d-2}^{2}\right) ,  \notag \\
&&\phi ^{i}=\phi ^{i}(r),\qquad A_{t}^{I}=A_{t}^{I}(r).
\label{eq:bhans}
\end{eqnarray}
Anticipating the explicit solutions of interest, we have
specialized to a black hole ansatz where the $g_{tt}$ and $g_{rr}$ warp
factors are appropriately related.  Doing so simplifies some of the
intermediate expressions below.  However this condition will be relaxed
when exploring thermodynamic considerations more generally in
section~\ref{sec:thermo}.

Before proceeding, note that the $(d-2)$-sphere may be parametrized as
\begin{equation}
d\Omega _{d-2}^{2}=d\psi ^{2}+\sin ^{2}\psi d\Omega _{d-3}^{2},
\end{equation}
in which case the $R_{\psi\psi}$ component of the Einstein equation,
(\ref{eq:eins}), takes the form
\begin{equation}
2R_{\psi }^{\psi }=-{\textstyle\frac{1}{2(d-2)}}G_{IJ}F_{\mu \nu }^{I}F^{\mu
\nu \,J}+{\textstyle\frac{2}{d-2}}V.
\label{eq:rpsipsi}
\end{equation}
This expression will prove useful below when evaluating the on-shell action.

\subsection{Stationary $R$-charged black holes}

Although the explicit form of the matter sector depends on the theory
of interest, the stationary $R$-charged black holes share a common
gravitational description.  In particular, the metric of (\ref{eq:bhans})
has the form
\begin{equation}
ds^2 = -{\cal H}(r)^{-(d-3)/(d-2)}f(r)\,dt^2+{\cal H}(r)^{1/(d-2)}
\left(\fft{dr^2}{f(r)}+r^2d\Omega_{d-2}^2\right),
\label{eq:bhmet}
\end{equation}
where
\begin{equation}
f(r)=1-\frac{\mu }{r^{d-3}}+g^{2}r^{2}\mathcal{H}(r).
\end{equation}
The function ${\cal H}(r)$ remains to be determined via the equations
of motion, and will be influenced by the set of matter fields and
charges that are turned on.  Nevertheless, in general ${\cal H}(r)$ admits
an expansion in inverse powers of $r$:
\begin{equation}
{\cal H}(r)=\prod_i H_i(r)=1+\fft{\alpha_1}{r^{d-3}}+\fft{\alpha_2}{r^{2(d-3)}}
+\fft{\alpha_3}{r^{3(d-3)}} +\cdots.
\label{eq:calhexp}
\end{equation}
For the solutions considered below
the function ${\cal H}(r)$ may be given explicitly as a product of harmonic
functions:
\begin{equation} 
 {\cal H}(r)=\prod_i H_i(r) = \prod_i \left(1+\fft{q_i}{r^{d-3}}\right). 
\end{equation}
In this case, the expansion coefficients in (\ref{eq:calhexp}) are
related to the charges $q_i$ according to
\begin{equation}
\alpha_1=\sum_iq_i,\qquad\alpha_2=\sum_{i<j}q_iq_j,\qquad
\alpha_3=\sum_{i<j<k}q_iq_jq_k,\qquad\hbox{\it etc.}.
\end{equation}
Note that, in the notation of (\ref{eq:bhans}), the warp factor
$B(r)$ is given simply by $B(r)=\fft1{2(d-2)}\log{\cal H}(r)$.  We will
examine these black holes in more detail in section~\ref{sec:bh4567}.
However, we first turn to the evaluation of the on-shell action,
corresponding to the thermodynamic potential $\Omega$.

\subsection{The regulated action and energy}

We now proceed to evaluate the on-shell action for spherically
symmetric configurations of the form (\ref{eq:bhans}). It is well
known that the action diverges due to the behavior of the metric
and matter fields near the boundary of an asymptotically $AdS_d$
spacetime. Anticipating these infrared divergences, a natural (but
non-covariant) way of regulating the calculation is to `cut-off'
the spacetime at a large but finite value of the $AdS$ radial
coordinate, $r=r_0$.  The result is a truncated spacetime, which
we denote $\MM_0$, whose `boundary' $\dM_0$ is located at the
cut-off.

We may now calculate the regulated action on the truncated spacetime.
To do so, we consider the bulk and boundary terms separately.  For the
bulk contribution, we first take the trace of the Einstein equation and
substitute it into (\ref{eq:bulka}) to obtain
\begin{equation}
I_{\mathrm{bulk}}=-{\frac{1}{16\pi G_{d}}}\int_{\MM_0} d^{d}x\sqrt{-g}\,
\left[-\fft1{2(d-2)}G_{IJ}F_{\mu\nu}^IF^{\mu\nu\,J}+\fft2{d-2}V\right].
\label{eq:ibgen}
\end{equation}
Using the equation of motion (\ref{eq:rpsipsi}), this expression may
be rewritten as
\begin{equation}
I_{\mathrm{bulk}}=-\fft1{8\pi G_{d}}\int_{\MM_0} d^{d}x\sqrt{-g}R_\psi^\psi.
\end{equation}
This may now be evaluated by explicit computation of $R_{\psi\psi}$ for the
black hole metric (\ref{eq:bhans}).  The result turns out to be a total
derivative
\begin{equation}
\sqrt{-g}R^\psi_\psi=-\fft{d}{dr}(r^{d-2}f(r)B'(r)+r^{d-3}(f(r)-1)).
\end{equation}
Hence
\begin{eqnarray}
I_{\rm bulk} &=&\frac{\beta\omega_{d-2}}{8\pi G_d}\left[ r^{d-2}f(r)B^{\prime
}(r)+r^{d-3}(f(r)-1)\right] _{r_{+}}^{r_0}\nonumber\\
&=&\frac{\beta \omega _{d-2}}{8\pi G_{d}}\left( r_{0}^{d-2}f(r_0)B^{\prime
}(r_0)+r_{0}^{d-3}(f(r_0)-1)+r_{+}^{d-3}\right),
\end{eqnarray}
where $r_+$ is the location of the horizon, given by $f(r_+)=0$.  The
factor $\beta=2\pi/T$ is the periodicity along the (Euclidean) time
direction, and $\omega_{d-2}$ is the volume of the unit $(d-2)$-sphere.

Turning to the Gibbons-Hawking surface term, we start by noting that the
unit normal in the $r$ direction is given by 
$n^{r}=e^{-B(r)}f(r)^{{\frac{1}{2}}}$.
Using the definition \eqref{ext_curv_defn} the components of the extrinsic
curvature tensor are:
\begin{eqnarray}
\Theta_{tt} &=& -h_{tt}e^{-B(r)}f(r)^{\fft12}\left(-(d-3)B'(r)
+\fft{f'(r)}{2f(r)}\right), \nonumber\\
\Theta_{\alpha\beta} &=& -h_{\alpha\beta}e^{-B(r)}f(r)^{\fft12}
\left(B'(r)+\fft1r\right),
\label{eq:thetas}
\end{eqnarray}
where indices $\alpha,\beta$ denote coordinates on the $(d-2)$-sphere. The
trace of the extrinsic curvature is then given by:
\begin{equation}
\Theta =-e^{-B(r)}f(r)^{{\frac{1}{2}}}\left(B'(r)+\frac{f'(r)}{2f(r)}
+\frac{d-2}{r}\right).
\end{equation}
The Gibbons-Hawking term, evaluated at the boundary of the regulated spacetime, is:
\begin{equation}
I_{\mathrm{GH}}=-\frac{\beta \omega _{d-2}}{8\pi G_{d}}\left(
r_{0}^{d-2}f(r_0)B^{\prime }(r_0)+{\frac{1}{2}}r_{0}^{d-2}f^{\prime }(r_0)
+(d-2)r_{0}^{d-3}f(r_0)\right).
\end{equation}
Assembling these terms, the regulated value of the on-shell action
(\ref{eq:bulka}) is given by
\begin{equation}
I_{\rm reg} = I_{\rm bulk}+I_{\rm GH}=\frac{\beta \omega_{d-2}}{8\pi G_{d}}
\left( -(d-3)r_{0}^{d-3}f(r_0)
-{\frac{1}{2}}r_{0}^{d-2}f^{\prime }(r_0)-r_{0}^{d-3}+r_{+}^{d-3}\right) .
\label{eq:unreg}
\end{equation}
This expression diverges as the cut-off is removed, $r_0\rightarrow\infty$,
and must be renormalized by an appropriate counterterm prescription.

Before addressing the counterterms, however, we first derive an expression
for the unrenormalized ADM energy.  To do so, we start with the unrenormalized
boundary stress tensor, given by
\begin{equation}
T^{ab}=\fft2{\sqrt{-h}}\fft{\delta I}{\delta h_{ab}}
=-\fft1{8\pi G_d}(\Theta^{ab}-\Theta h^{ab}).
\label{eq:bst}
\end{equation}
Making use of (\ref{eq:thetas}), the time-time component of the
stress tensor has the form
\begin{equation}
\sqrt{-h}T_{tt}=-\fft1{8\pi G_d}h_{tt}(d-2)(r^{d-2}f(r)B'(r)+r^{d-3}f(r))
\end{equation}
so that
\begin{equation}
E_{\rm reg}=-\fft{\omega_{d-2}}{8\pi G_d}(d-2)(r_{0}^{d-2}f(r_0)B'(r_0)
+r^{d-3}f(r_0))
\label{eq:eunreg}
\end{equation}
While this ADM energy also diverges as the cut-off is removed,
 the difference $(I_{\rm reg}-\beta E_{\rm reg})$
is finite in this limit.  In other words, the difference between the
thermodynamic potential and the energy is {\it a priori} finite, and
does not need renormalization.  Nevertheless, one is often interested in
understanding the energy of the system on its own, and in this case
a proper choice of counterterms must be made.  We now turn to a
Hamilton-Jacobi analysis in order to fix the counterterm action.

\section{Hamilton-Jacobi counterterms}

\label{sec:hjcounterterms}

As we have seen above, the on-shell action for gravity on an asymptotically
AdS spacetime typically contains infrared divergences related to the
behavior of the  metric (and any other fields) near the boundary.  We now
review the calculation of boundary counterterms and demonstrate that the
Hamilton-Jacobi method generates appropriate counterterms for canceling
all power-law divergences in the on-shell action.

In order to facilitate the Hamiltonian
analysis it is convenient to foliate this spacetime with constant $r$
hypersurfaces, orthogonal to a spacelike unit normal $n^\mu$.
The hypersurface defined by the cut-off $r=r_0$ can be thought of as the
`boundary' of the regulated spacetime, with
\begin{eqnarray}
  \lim_{r_0 \to \infty} \dM_0  & = & \dM.
\end{eqnarray}
Using the Gauss-Codacci equations, the action (\ref{eq:bulka}) can be
rewritten in terms of the intrinsic curvature ${\cal R}$ of the hypersurfaces
and the extrinsic curvature $\Theta_{ab}$ describing their embedding
in $\MM_0$. Note that, now that we have fixed the normal to point in the $r$
direction, we will use indices $a,b,\ldots$ for tensors defined on the
constant $r$ hypersurfaces of the foliation.  The regulated action is then
given by:
\begin{eqnarray}
\label{reg_action}
  I &=& -\fft1{16\pi G_d}\int_{\MM_0}\nts d^{\,d}x \sqrt{-g} \,
   \bigl[{\cal R}+\Theta^2-\Theta^{ab}\Theta_{ab}
   -\ft12\,{\cal G}_{ij}(\phi)n^{\mu}\partial_{\mu}\phi^i n^{\nu}
   \partial_\nu\phi^j \nonumber\\
  & &\kern4em -\ft12\,{\cal G}_{ij}(\phi)h^{ab}\partial_{a}\phi^i
   \partial_{b}\phi^j
   -\ft14\,G_{IJ}(\phi)\, F^{ab\, I} F_{ab}^{J} \nonumber \\  
  & &\kern4em -\ft12\,G_{IJ}(\phi)h^{ab}\, n^{\mu} F_{\mu a}^{I} n^{\nu}
   F_{\nu b}^{J} - V(\vf)\bigr].
\end{eqnarray}
The action \eqref{reg_action} is explicitly quadratic in first derivatives
of the fields $\vf^i$, $A_{\mu}^{\,I}$, and $\tg_{\mu\nu}$.  Taking into
account the holographic principle of flows in the radial direction, we
define conjugate momenta and the Hamiltonian with respect to the AdS radial
coordinate $r$, as opposed to the usual choice of a time coordinate.
In this case, the momenta conjugate to these fields are given by:
\begin{eqnarray}
\label{momenta}
 \pi_i &=& \fft1{16\pi G_d}{\cal G}_{ij}(\phi) \, n^{\mu} \partial_{\mu}
   \vf^{j},\nonumber\\
 \pi^a_{\,I} & = & \fft1{16\pi G_d}G_{IJ}(\vf)\,\tg^{ab}
   n^{\mu} F_{\mu b}^{\,J}, \nonumber\\
 \pi^{ab} & = & \frac{1}{16\pi G_d}\,\left( \tg^{ab} \,\Theta
    - \Theta^{ab}\right).
\end{eqnarray}
Using these momenta, the Hamiltonian density obtained from
\eqref{reg_action} is:
\begin{eqnarray}
\label{Hamiltonian_density}
  \HH &=& 16\pi G_d \left(\frac{1}{2}\,{\cal G}_{ij}(\phi)\pi^i \pi^j
      + \pi^{ab}\pi_{ab} -\frac{1}{d-2}\,\pi^{a}_{\,a} \pi^{b}_{\,b}
       + \frac{1}{2}\,G_{IJ}\,\tg^{ab}
      \pi^I_a \pi^J_b\right)\nonumber\\[2pt]    
    &&+\frac1{16\pi G_d}\left(\, {\cal R} - \fft12\,
      {\cal G}_{ij}(\phi) h^{ab} \partial_a\vf^i\partial_b\vf^{j}
      -V(\vf)- \frac{1}{4} G_{IJ} F_{ab}^{\,I}F^{ab\,J}\right)\nonumber\\[2pt]
    &&+G_{IJ} \tg^{ab} \pi^I_a n^\mu\partial_bA_\mu^{\,J}.
\end{eqnarray}
Diffeomorphism invariance of the theory constrains the Hamiltonian
(and other generators of coordinate transformations) to vanish.
in other words,
\begin{equation}
\label{Ham_Constraint}
  \HH[\pi_i,\vf^i,\pi^a_{\,I},A_a^{\,I},\pi^{ab},\tg_{ab}]=0.
\end{equation}
To obtain the Hamilton-Jacobi equation we must rewrite the Hamiltonian
constraint in terms of functional derivatives of the on-shell action. The
on-shell action is a functional of the bulk fields evaluated at the
boundary $\dM_0$. According to Hamilton-Jacobi theory the variational
derivative of the on-shell action with respect to a field's boundary value
gives the momenta conjugate to that field, evaluated at $\dM_0$. Thus,
the momenta \eqref{momenta} can be written as functional derivatives of
the on-shell action:
\begin{equation}
\label{HJmomenta}
  \pi_i = \frac{1}{\sqrt{-\tg}} \, \frac{\delta I}{\delta \vf^i},
	\qquad
  \pi^a_{\,I} = \frac{1}{\sqrt{-\tg}} \, \frac{\delta I}{\delta A_a^{\,I}},
	\qquad
  \pi^{ab} = \frac{1}{\sqrt{-\tg}} \, \frac{\delta I}
    {\delta \tg_{ab}},
\end{equation}
where the fields in \eqref{HJmomenta} are evaluated at $r_0$.
Finally, replacing the momenta appearing in the Hamiltonian
with functional derivatives of the on-shell action, we obtain the
Hamilton-Jacobi equation:
\begin{equation}
\label{HJ_eqn}
   \HH\left[\frac{\delta I}{\delta \vf^i},\vf^i,
      \frac{\delta I}{\delta A_a^{\,I}},A_a^{\,I},
      \frac{\delta I}{\delta \tg_{ab}},\tg_{ab} \right] = 0.
\end{equation}
The Hamilton-Jacobi equation is a functional differential equation for
the on-shell action in terms of the boundary values of the bulk fields.

\subsection{Derivation of the counterterm action}

Using the Hamilton-Jacobi equation, we can obtain a set of counterterms
that will remove power-law divergences from the on-shell action. We
first write the regulated on-shell action as:
\begin{equation}
\label{decompose_action}
    I_{reg} = \Gamma - I_{ct}
\end{equation}
The first term, $\Gamma$, represents the part of the action which is finite%
\footnote{In general $\Gamma$ might contain logarithmic divergences. These
divergences, which are related to the Weyl anomaly in the dual field
theory, can be addressed using the Hamilton-Jacobi approach. However, the
gauged supergravity solutions we consider are free of such divergences.}
upon removing the cut-off. The second term, $I_{ct}$, represents the
power-law divergences appearing in the action. The terms appearing in $I_{ct}$
are conveniently organized in terms of an inverse metric expansion, as
described in \cite{Martelli:2002sp}. A sufficient counterterm action for
the gauged supergravity black hole solutions we consider is given by:
\begin{equation}
\label{ctaction}
  I_{ct} = \frac 1{8\pi G_d}\int_{\dM_0} \nts d^{d-1}x \sqrt{-\tg}
    \left(W(\vf) + C(\vf) \,{\cal R} + D(\vf) {\cal R}^2 +
    E(\vf) {\cal R}^{ab} {\cal R}_{ab}\right).
\end{equation}
The first two terms contain the divergences that appear in four and five
dimensions, while in six and seven dimensions it is necessary to include
the remaining terms. In constructing this action we have discarded a
number of possible gradient counterterms of the form $M_{ij}(\phi)
\partial_a\phi^i\partial^a\phi^j$, {\it etc.}, because the scalar fields
only depend on the AdS radial coordinate $r$.  In addition, since the
counterterm action should respect any residual bulk symmetries,
the $U(1)$ gauge fields should only appear in terms of gauge-invariant
field strengths $F_{ab}^{\,I}$. These terms do not contribute to
\eqref{ctaction} for the electrically charged configurations given by the
ansatz \eqref{eq:bhans}.
It is important to note, however, that if one wishes to study fluctuations
around the black hole backgrounds then such counterterms must be included
in \eqref{ctaction}, since the fluctuations may depend on the transverse
coordinates. For such cases, the counterterm action \eqref{ctaction}
alone is not sufficient for calculations of correlators in the field
theory duals of these solutions.

The momenta can be decomposed into contributions from the terms in
\eqref{decompose_action}, schematically of the form:
\begin{equation}
    \pi = \pi_{\Gamma}-P.
\end{equation}
The contributions $P$ due to the counterterm action are given by functional
derivatives of \eqref{ctaction} with respect to the fields on $\dM_0$:
\begin{eqnarray}
  P_i & = & \frac{1}{\sqrt{-\tg}}~\frac{\delta I_{ct}}{\delta \vf^i}
	\nonumber\\
      & = & \frac 1{8\pi G_d}\left(\frac{\partial \, W}{\partial \vf^i}
	+ \frac{\partial \, C}{\partial \vf^i}\, {\cal R}
	+ \frac{\partial \, D}{\partial \vf^i}\, {\cal R}^2
	+ \frac{\partial \, E}{\partial \vf^i}\, {\cal R}^{ab}
	{\cal R}_{ab}\right),\nonumber\\
  P^{ab} &=& \frac{1}{\sqrt{-\tg}}~\frac{\delta I_{ct}}
	{\delta \tg_{ab}}\nonumber\\
       &=&\frac 1{8\pi G_d}\Bigl( \frac{1}{2}\,\tg^{ab} \, W
	-C\, {\cal G}^{ab} + \frac{1}{2}\,\tg^{ab}
	D \, {\cal R}^2 + \frac{1}{2}\,\tg^{ab} E \,
	{\cal R}^{cd} {\cal R}_{cd}\nonumber\\
  & &\qquad -2 D\, {\cal R} {\cal R}^{ab} + 2 E
	{\cal R}^{dba}{}_c~{\cal R}^c{}_d
	- E\,\nabla_c\nabla^c {\cal R}^{ab}\Bigr).
\end{eqnarray}
The term ${\cal G}^{ab}$ appearing in the expression for $P^{ab}$ is
the boundary Einstein tensor, given by:
\begin{equation}
   {\cal G}_{ab} = \RR_{ab}- \ft12\,h_{ab} \RR
\end{equation}
The counterterms $W(\phi)$, $C(\phi),\ldots$ are now determined by
substituting these momenta into the Hamilton-Jacobi equation
\eqref{HJ_eqn} and solving it order-by-order in the expansion
\eqref{ctaction}. We denote the various terms in the Hamiltonian by
\begin{equation}
\label{H_expansion}
 \HH = \HH_{(0)} + \HH_{(1)} + \HH_{(2)} + \ldots + \HH_{\Gamma}.
\end{equation}
The terms $\HH_{(i)}$ represent contributions from $I_{\rm ct}$, with the
index counting the number of inverse metrics appearing in that term. For
the backgrounds we are interested in this is an adequate measure of the
degree of divergence these terms represent. Evaluating these terms leads
to differential equations for the functions appearing in \eqref{ctaction}.
The most illuminating of these is the equation for $W(\vf)$ that comes
from the term $\HH_{(0)}$ in the Hamiltonian constraint:
\begin{eqnarray}
\label{H0eqn}
  \HH_{(0)} &=& \frac1{16\pi G_d}\left(2\,{\cal G}^{ij}(\phi)
     \,\frac{\partial \, W}{\partial \vf^i}
     \frac{\partial \, W}{\partial \vf^j} -
     \, \frac{d-1}{d-2} W^2 - V\right).
\end{eqnarray}
Setting $\HH_{(0)} = 0$ recasts \eqref{H0eqn} as the familiar
relation for the potential $V(\vf)$ in terms of the superpotential
$W(\vf)$:
\begin{equation}
\label{superpotential_eqn}
V = 2{\cal G}^{ij}(\phi)\fft{\partial W}{\partial\phi^i}
\fft{\partial W}{\partial\phi^j}-\fft{d-1}{d-2}W^2.
\end{equation}
The conclusion is that the leading term in the counterterm action
\eqref{ctaction} is simply proportional to the superpotential $W(\vf)$
\cite{deBoer:1999xf}.

We obtain similar equations for the functions $C(\vf)$, $D(\vf)$, and
$E(\vf)$ by evaluating the remaining terms in \eqref{H_expansion}. The
equation derived from $\HH_{(1)}=0$ determines $C(\vf)$ in terms
of $W(\vf)$:
\begin{equation}
\label{C_eqn}
  \frac{1}{2}+2{\cal G}^{ij}(\phi)~\frac{\partial \, W}{\partial \vf^i}
    ~\frac{\partial \, C}{\partial \vf^j} - \frac{d-3}{d-2}\,C\,W = 0.
\end{equation}
The counterterms $W(\vf)$ and $C(\vf)$, determined by equations
\eqref{superpotential_eqn} and \eqref{C_eqn}, completely characterize
the power-law divergences in four and five dimensions. For the
six and seven dimensional supergravities there are two additional
counterterms whose coefficients $D(\vf)$ and $E(\vf)$ are determined
by two equations obtained from functionally independent terms in the
equation $\HH_{(2)}=0$:
\begin{eqnarray}
 -{\cal G}^{ij}(\phi) \, \frac{\partial \, C}{\partial \vf^i}
   \frac{\partial \, C}{\partial \vf^j}
   -2{\cal G}^{ij}(\phi)\,\frac{\partial \, W}{\partial \vf^i}
   \frac{\partial \, D}{\partial \vf^j} +  \frac{d-5}{d-2}
   \,D\,W + \frac{d-1}{2(d-2)}\,C^2 & = & 0,\nonumber\\
 -2 {\cal G}^{ij}(\phi)\, \frac{\partial W}{\partial \vf^i}
    \frac{\partial \, E}{\partial \vf^j} + \frac{d-5}{d-2}
   \,E\,W-2 C^2 & = & 0.
\label{DE_eqn}
\end{eqnarray}
In five dimensions, where $D(\vf)$ and $E(\vf)$ are not included in the
counterterm action \eqref{ctaction}, $\HH_{(2)}$ actually represents
a potentially non-vanishing term in the expansion \eqref{H_expansion}
for $\HH$:
\begin{equation}
\label{extra_term}
\HH_{(2)} = \frac{1}{8\pi G_d}\left( \GG^{ij}(\vf)\,
    \frac{\partial \, C}{\partial \vf^i}
    \frac{\partial \, C}{\partial \vf^j}~{\cal R}^2
    + 2 \,C^2 ({\cal R}^{ab}{\cal R}_{ab}-\ft13{\cal R}^2)
    \right).
\end{equation}
In principle such a term might signal the presence of a logarithmic
divergence in the on-shell action, corresponding to a Weyl anomaly in
the dual field theory. However, for the solutions we are interested in
the terms appearing in \eqref{extra_term} either vanish due to the $S^1
\times S^{d-2}$ topology of the boundary, or vanish sufficiently rapidly
near the boundary so as to not contribute any additional divergences to
the effective action.

While we have shown, in equation \eqref{superpotential_eqn}, that the
leading counterterm is simply the superpotential $W(\vf)$, we have
not provided explicit solutions for the remaining terms. 
For the gauged supergravity solutions we are interested in
it is sufficient to solve for the functions $C(\vf)$, $D(\vf)$,
and $E(\vf)$ as a power series in $\vf^i$, out to order $\OO(\vf^2)$.
However, rather than writing general solutions, which would depend on the
choice of basis for the gauged supergravity scalars, we will specialize
to an appropriate expansion for each of the $d$-dimensional black
holes that we consider in the next section. Finally, it should be noted
that the functions $C(\vf)$, $D(\vf)$, and $E(\vf)$ can be written in
terms of integrals of the superpotential and its derivatives, but these
expressions are not particularly illuminating.

\subsection{Counterterm renormalization of the energy}

Since the counterterm $W(\phi)$ is simply related to the superpotential
according to (\ref{superpotential_eqn}), its form is already determined.
For the remaining counterterms, $C(\phi)$, $D(\phi)$ and $E(\phi)$, their
solutions as power series expressions may be motivated by noting that the
large $r$ asymptotics of the black hole solution (\ref{eq:bhans}) generically
has the form
\begin{equation}
f(r)\sim g^2 r^2,\qquad B(r)\sim \fft1{r^{d-3}},\qquad
\phi^i(r)\sim \fft1{r^{d-3}},\qquad A_t^I(r)\sim \fft1{r^{d-3}}.
\end{equation}
To cancel divergences, and to provide possibly finite counterterms, the
series solution to $C(\phi)$ must be determined to ${\cal O}(1/r^{d-3})$
while the series solutions to $D(\phi)$ and $E(\phi)$ must be determined
to ${\cal O}(1/r^{d-5})$.  As a result, only the leading terms will be
important
\begin{eqnarray}
C(\phi)&=&c_0+c_i\phi^i+\hbox{unimportant},\nonumber\\
D(\phi)&=&d_0+\hbox{unimportant},\nonumber\\
E(\phi)&=&e_0+\hbox{unimportant}.
\end{eqnarray}

Assuming, from symmetry, that the linear term $c_i$ is absent in $C(\phi)$,
we only need to compute the constant pieces $c_0$, $d_0$ and $e_0$ from
(\ref{C_eqn}) and (\ref{DE_eqn}).  As a result, we find that the relevant
contribution of the counterterm action has the form
\begin{eqnarray}
I_{\rm ct}&=&\fft1{8\pi G_d}\int d^{d-1}x\sqrt{-h}\biggl(W(\phi)
+\fft{1}{2(d-3)g}{\cal R}\nonumber\\
&&\qquad +\fft{1}{2(d-5)(d-3)^2g^3}\left({\cal R}_{ab}{\cal R}^{ab}
-\fft{d-1}{4(d-2)}{\cal R}^2\right)+\cdots\biggr).
\label{eq:ict}
\end{eqnarray}
This can be compared with similar expressions for pure gravitational
backgrounds, as found in \cite{Balasubramanian:1999re,Emparan:1999pm}.
Note that $g$ is also the inverse of the $AdS$ length scale, $\ell^{-1}$,
which is given in terms of the constant term $V_0$ in the scalar potential
$V(\phi)$ by:
\begin{equation}
  \ell = \sqrt{-\frac{(d-1)(d-2)}{V_0}}.
\end{equation}
Corresponding to the counterterms in (\ref{eq:ict}), the regulated boundary
stress tensor picks up an additional contribution
\begin{eqnarray}
T_{\rm ct}^{ab}&=&\fft1{8\pi G_d}\biggl(h^{ab}W(\phi)
-\fft{1}{2(d-3)g}(2{\cal R}^{ab}-{\cal R}h^{ab})
+\fft{1}{2(d-5)(d-3)^2g^3}\biggl(
4{\cal R}^{acbd}{\cal R}_{cd}
\nonumber\\
&& \qquad -\fft{d-1}{d-2}{\cal R}^{ab}{\cal R}  + h^{ab}
\Bigl({\cal R}_{cd}{\cal R}^{cd}-\fft{d-1}{4(d-2)}{\cal R}^2
\Bigr)\biggr)\biggr).
\end{eqnarray}
Some terms proportional to derivatives of ${\cal R}$ along the boundary
have been omitted, as they vanish for the spherically symmetric solutions
of interest.

For black hole metrics of the form (\ref{eq:bhans}), the boundary Ricci
tensor is given by
\begin{equation}
{\cal R}_{tt}=0,\qquad
{\cal R}_{\alpha\beta}=(d-3)h_{\alpha\beta}e^{-2B}r^{-2}.
\end{equation}
Thus the counterterm action and contribution to the energy may be expressed as
\begin{equation}
E_{\rm ct}=\fft{I_{\rm ct}}{\beta}
=\fft{\omega_{d-2}}{8\pi G_d}e^Bf^{\fft12}\left(r^{d-2}W(\phi)
+\fft{(d-2)}{2\,g}e^{-2B}r^{d-4}-\fft{(d-2)}{8\,g^3}e^{-4B}r^{d-6}
+\cdots\right),
\label{eq:bhct}
\end{equation}
where
\begin{equation}
E_{\rm ct} = \omega_{d-2}\sqrt{-h}\, h^{tt} T_{tt}^{\rm ct}
\end{equation}
is the counterterm contribution to the ADM energy.  The relation $I_{\rm
ct}=\beta E_{\rm ct}$ demonstrates that, while the counterterms are
necessary to render both the action and the energy 
finite, the validity of the thermodynamic relation $\Omega = E-TS-Q_I\Phi^I$
is unaffected by any finite shift in the counterterm action.

\section{The renormalized action and mass}
\label{sec:bh4567}

Given the regulated action (\ref{eq:unreg}) and energy (\ref{eq:eunreg}),
as well as the corresponding counterterm expressions (\ref{eq:ict}) and
(\ref{eq:bhct}), we are now in a position to examine the various
$R$-charged black holes. In each case we calculate the renormalized
action $\Gamma$ and energy $E_{ren}$, and show that they are finite in
the $r_0 \rightarrow \infty$ limit.

\subsection{$D=4$ black holes}

In four dimensions, the ${\cal N}=2$ truncation of gauged ${\cal N}=8$
supergravity yields a system with three complex scalars and four $U(1)$
gauge fields.  For simplicity, we consider a truncation of the scalar
sector by setting the axionic components to zero.  While this is in
principle an inconsistent truncation, this is nevertheless a valid
procedure when applied to the non-rotating electrically charged black
holes.  In this case, the three dilatonic scalars may be parametrized
by a constrained set of real fields $X_i$ satisfying $X_1X_2X_3X_4=1$.
The potential and superpotential are then given by
\begin{equation}
V=-g^2\sum_{i<j}X_iX_j,\qquad W=\ft12g\sum_iX_i.
\label{eq:4pot}
\end{equation}

In addition to the metric, (\ref{eq:bhmet}), the four-charge
black holes have gauge potentials and scalars given by
\cite{Duff:1999gh,Sabra:1999ux,Cvetic:1999xp}
\begin{equation}
A_{(1)}^i=\sqrt{\fft{q_i+\mu}{q_i}}\left(1-\fft1{H_i}\right)dt,\qquad
X_i =\fft{{\cal H}^{1/4}}{H_i}.
\end{equation}
As a result, the regulated action integral, (\ref{eq:unreg}), becomes
\begin{equation}
I_{\rm reg}=\frac{\beta \omega _{2}}{8\pi G_4}
\left(-2g^2r_0^3-\fft32g^2\alpha_1r_0^2-(2+g^2\alpha_2)r_0+\fft12\mu
-\fft12g^2\alpha_3+r_+\right)
\end{equation}
Note that the first three terms are divergent as $r_0 \rightarrow \infty$.
Of course, the boundary counterterm remains to be evaluated.  To do so,
we simply insert the form of the superpotential, given by (\ref{eq:4pot}),
into (\ref{eq:bhct}) to obtain
\begin{equation}
\beta E_{\rm ct}=I_{\rm ct}=\fft{\beta\omega_2}{8\pi G_4}\left(2g^2r_0^3
+\fft32g^2\alpha_1r_0^2+(2+g^2\alpha_2)r_0-\mu+\fft12g^2\alpha_3\right).
\end{equation}
We now see explicitly that the divergent terms in the regulated action
are canceled by the counterterms.  Furthermore, the nonlinear charge
term, proportional to $\alpha_3$, vanishes in the renormalized action
\begin{equation}
\Gamma=I_{\rm reg}+I_{\rm ct}=\fft{\beta\omega_2}{8\pi G_4}
\left(-\fft12\mu+r_+\right).
\end{equation}
Turning to the ADM energy, we first evaluate (\ref{eq:eunreg}) in four
dimensions to obtain
\begin{equation}
E_{\rm reg}=\fft{\omega_2}{8\pi G_4}\left(-2g^2r_0^3-\fft32g^2\alpha_1r_0^2
-(2+g^2\alpha_2)r_0+2\mu+\fft12\alpha_1-\fft12g^2\alpha_3\right).
\end{equation}
Combining this with $E_{\rm ct}$ yields the linear mass/charge relation
\begin{equation}
E_{\rm ren}=\fft{\omega_2}{8\pi G_4}\left(\mu+\fft12\alpha_1\right)
=\fft{\omega_2}{8\pi G_4}\left(\mu+\fft12(q_1+q_2+q_3+q_4)\right).
\end{equation}

\subsection{$D=5$ black holes}

As in the ungauged case, gauged $D=5$, ${\cal N}=2$ supergravity 
coupled to an arbitrary number of vector multiplets has a natural
description in terms of special geometry.  Here we consider only the
particular case of the STU model, corresponding to the $U(1)^3$
truncation of maximal gauged supergravity.  The black holes in this model,
which may carry up to three charges, have been well studied
\cite{Cvetic:1999ne}.

The counterterm renormalization prescription for black holes in the
STU model was recently examined in \cite{Buchel:2003re} for single-charge
black holes and in \cite{Liu:2004it} for three-charge black holes.
In this model, the potential and superpotential are given by
\begin{equation}
V=-4g^2\sum_{i<j}X_iX_j=-4g^2\sum_i\fft1{X_i},\qquad W=g\sum_iX_i,
\end{equation}
where the two real scalars are encoded in the constrained fields
$X_1X_2X_3=1$.  Furthermore, the gauge potentials and scalars have the
form \cite{Behrndt:1998ns,Behrndt:1998jd}
\begin{equation}
A_{(1)}^i=\sqrt{\fft{q_i+\mu}{q_i}}\left(1-\fft1{H_i}\right)dt,\qquad
X_i =\fft{{\cal H}^{1/3}}{H_i}.
\end{equation}

Working out the regulated action and energy, we find the similar expressions
\begin{eqnarray}
I_{\rm reg}&=&\fft{\beta\omega_3}{8\pi G_5}
\left(-3g^2r_0^4-(3+2g^2\alpha_1)r_0^2+\mu-g^2\alpha_2+r_+^2\right),\nonumber\\
E_{\rm reg}&=&\fft{\omega_3}{8\pi G_5}
\left(-3g^2r_0^4-(3+2g^2\alpha_1)r_0^2+3\mu+\alpha_1-g^2\alpha_2\right).
\end{eqnarray}
The divergences are renormalized by the counterterm action
%
\begin{equation}
\beta E_{\rm ct}=I_{\rm ct}=\fft{\beta\omega_3}{8\pi G_5}
\left(3g^2r_0^4+(3+2g^2\alpha_1)r_0^2-\fft32\mu+\fft3{8g^2}+g^2\alpha_2\right).
\end{equation}
Consequently, we find the familiar results
\begin{equation}
\Gamma=\fft{\beta\omega_3}{8\pi G_5}
\left(-\fft12\mu+r_+^2+\fft3{8g^2}\right),
\end{equation}
and
\begin{equation}
E_{\rm ren}=\fft{\omega_3}{8\pi G_5}\left(\fft32\mu+\alpha_1+\fft3{8g^2}\right)
=\fft{\omega_3}{8\pi G_5}\left(\fft32\mu+q_1+q_2+q_3+\fft3{8g^2}\right).
\end{equation}

\subsection{$D=6$ black holes}

In six dimensions, the gauged ${\cal N}=(1,1)$ supergravity admits two
inequivalent AdS vacua \cite{Romans:tw}, only one which is supersymmetric.
It is this one that we consider.  The bosonic components of the
supergravity multiplet consists of a graviton $g_{\mu\nu}$, antisymmetric
tensor $B_{\mu\nu}$, $SU(2)\times U(1)$ gauge fields $A_\mu^I$, ${\cal
A}_\mu$ and a dilaton $\phi$.  The potential and superpotential have
the form
\begin{equation}
V=-g^2\left(9X^2+\fft{12}{X^2}-\fft1{X^6}\right),\qquad
W=g\left(3X+\fft1{X^3}\right)
\end{equation}
where $X=e^{-\fft1{2\sqrt{2}}\phi}$.

The gauging of \cite{Romans:tw} which leads to an AdS$_6$ vacuum also
turns on a mass for the antisymmetric tensor.  More directly, the abelian
vector ${\cal A}_\mu$ is absorbed by $B_{\mu\nu}$ for mass generation.
Thus we only consider abelian black holes charged under the $U(1)$
subgroup of $SU(2)$.  The gauge potential and dilaton are given by
\begin{equation}
A_{(1)}^3=\sqrt{\fft{q+\mu}q}\left(1-\fft1H\right)dt,\qquad X=H^{-1/4}.
\end{equation}
Note, however, that ${\cal H}=H^2$, so that $\alpha_1=2q$ and $\alpha_2=q^2$.

The regulated six-dimensional action is
\begin{equation}
I_{\rm reg}=\frac{\beta \omega _{4}}{8\pi G_6}\left(-4g^2r_0^5-4r_0^{3}
-5g^{2}qr_0^{2}+\frac{3}{2}\mu+r_{+}^{3}\right),
\end{equation}
The regulated ADM energy is similarly
\begin{equation}
E_{\rm reg}=\fft{\omega_4}{8\pi G_6}\left(-4g^2r_0^5-4r_0^3-5g^2qr_0^2
+3q+4\mu\right).
\end{equation}
At the same time, evaluation of the boundary counterterm, (\ref{eq:bhct}),
yields
\begin{equation}
\beta E_{\rm ct}=I_{\rm ct}=\fft{\beta\omega_4}{8\pi G_6}
\left(4g^2r_0^5+4r_0^3+5g^2qr_0^2-2\mu\right)
\end{equation}
This is the first case when the curvature-squared counterterms turn out
to be important.  We end up with simple expressions for the regulated
action and ADM energy
\begin{eqnarray}
\Gamma&=&\fft{\beta\omega_4}{8\pi G_6}
\left(-\fft12\mu+r_+^3\right),\nonumber\\
E_{\rm ren}&=&\fft{\omega_4}{8\pi G_6}\left(2\mu+3q\right).
\end{eqnarray}
Note the absence of any Casimir energy for the odd-dimensional boundary
theory.

\subsection{$D=7$ black holes}

Maximal gauged supergravity in seven dimensions involves the gauging of
an $SO(5)$ $R$-symmetry, as can be deduced from the $S^4$ reduction of
eleven-dimensional supergravity.  This can be truncated to half-maximal
supergravity (with $SU(2)$ gauging) coupled to an abelian vector
multiplet.  For simplicity, however, we consider a further truncation to
two abelian vectors and two scalars.  In general, this is no longer a
consistent supergravity theory.  However, it is consistent to consider
a subset of solutions, including the electrically charged black holes
of present interest.

Because of the slightly unusual nature of the truncated theory, the
potential has a more complicated structure \cite{Liu:1999ai,Cvetic:1999xp}
\begin{equation}
V=-2g^2\left(8X_1X_2+\fft4{X_1^2X_2}+\fft4{X_1X_2^2}-\fft1{X_1^4X_2^4}\right),
\end{equation}
where $X_1$ and $X_2$ are unconstrained fields.  In terms of canonically
normalized scalars, we may take the representation
\begin{equation}
X_1=e^{\fft1{\sqrt{10}}\varphi_1+\fft1{\sqrt{2}}\varphi_2},\qquad
X_2=e^{\fft1{\sqrt{10}}\varphi_1-\fft1{\sqrt{2}}\varphi_2}.
\end{equation}
The superpotential has the form
\begin{equation}
W=g\left(2X_1+2X_2+\fft1{X_1^2X_2^2}\right).
\end{equation}

For the $R$-charged black holes, the two gauge potentials and
two scalars are given in terms of the harmonic functions $H_i$ by
\cite{Liu:1999ai,Cvetic:1999xp}
\begin{equation}
A_{(1)}^i=\sqrt{\fft{q_i+\mu}{q_i}}\left(1-\fft1{H_i}\right)dt,\qquad
X_i=\fft{{\cal H}^{2/5}}{H_i}.
\end{equation}
This yields the expression for the superpotential
\begin{equation}
W=g{\cal H}^{2/5}\left(\fft2{H_1}+\fft2{H_2}+1\right).
\end{equation}
The resulting regulated on-shell action is
\begin{equation}
I_{\rm reg}=\fft{\beta\omega_5}{8\pi G_7}
\left(-5g^2r_0^6-5r_0^4-3g^2\alpha_1r_0^2+2\mu+r_+^4\right),
\end{equation}
and the regulated ADM energy is
\begin{equation}
E_{\rm reg}=\fft{\omega_5}{8\pi G_7}
\left(-5g^2r_0^6-5r_0^4-3g^2\alpha_1r_0^2+5\mu+2\alpha_1\right).
\end{equation}
Note that these expressions are already at most linear in the charges.

In six or higher dimensions, the asymptotic scalar behavior falls
off sufficiently rapidly so that the scalars do not contribute to the
boundary counterterm.  We find
\begin{equation}
\beta E_{\rm ct}=I_{\rm ct}=\fft{\beta\omega_5}{8\pi G_7}
\left(5g^2r_0^6+5r_0^4+3g^2\alpha_1r_0^2-\fft52\mu-\fft5{16g^4}\right),
\end{equation}
so that the renormalized values are
\begin{equation}
\Gamma=\fft{\beta\omega_5}{8\pi G_7}
\left(-\fft12\mu+r_+^4-\fft5{16g^4}\right),
\end{equation}
and
\begin{equation}
E_{\rm ren}=\fft{\omega_5}{8\pi G_7}
\left(\fft52\mu+2\alpha_1-\fft5{16g^4}\right).
\end{equation}
%

\section{Black hole energy and thermodynamics}
\label{sec:thermo}

In the previous sections we demonstrated explicitly that the on-shell
action and the ADM energy may be renormalized by introducing an appropriate
counterterm action given by a Hamilton-Jacobi analysis.  Turning to the
dual field theory, the renormalized on-shell action is to be identified
with the thermodynamic potential $\Omega$ according to $\Gamma=\beta\,\Omega$.
Likewise, the ADM energy $E_{\rm ren}$ ought to be identified with the
energy (including Casimir energy) of the field theory.

For backgrounds with non-trivial $R$-charge, the thermodynamic potential
may be related to the energy according to
\begin{equation}
\Omega=E-TS-\Phi^IQ_I,
\label{eq:thermo}
\end{equation}
where $Q_I$ are the set of conserved $R$-charges, and $\Phi^I$ are the
corresponding horizon values of the electric potential.  Here we prove
that the relation (\ref{eq:thermo}) is automatically satisfied for the
black hole solutions of the previous section.

We start with a static, stationary metric in $d$ dimensions, of the form
(\ref{eq:bhans})
\begin{equation}
ds_d^2=-e^{2A}f\,dt^2+e^{2B}\left(\fft{dr^2}{f}+r^2d\Omega_{d-2}^2\right).
\label{eq:ssmet}
\end{equation}
Note, however, that here we allow independent warp factors for the time
and space directions.
This choice of coordinates is specialized so that the boundary of AdS is
located at $r\to\infty$ and also so that $\partial/\partial t$ is a natural
time-like Killing vector.  We further assume that the matter sector
preserves the time translation symmetry, so that in particular all matter
fields are independent of $t$.

As in (\ref{eq:bulka}) the unrenormalized action integral is composed of
two pieces, the bulk integral and the surface term.  To evaluate the bulk
action we start with the expression (\ref{eq:ibgen}).  However, instead
of using the $R_{\psi\psi}$ component of the Einstein equation, we
substitute in the $R_{tt}$ component to rewrite the bulk action as
\begin{equation}
I_{\rm bulk}=-\fft1{8\pi G_d}\int_{\MM_0} d^dx\sqrt{-g}\left(R^t_t
-\fft12G_{IJ}F_{tr}^IF^{J\,tr}\right).
\label{eq:bulktt}
\end{equation}
We now show that this bulk integrand is in fact a total divergence.  First
note that, for the metric (\ref{eq:ssmet}), the $tt$ component of the
Ricci tensor may be written as
\begin{equation}
R^t_t=\fft1{\sqrt{-g}}\fft{d}{dr}\left(\sqrt{-h}\Theta^t_t\right),
\label{eq:rttval}
\end{equation}
which is already a total derivative.  For the gauge fields, on the other
hand, we recall that they satisfy the equation of motion (\ref{eq:eins})
so that
\begin{equation}
\partial_r(\sqrt{-g}G_{IJ}F^{J\,rt})=0.
\end{equation}
As a result, we may define the conserved charges
\begin{equation}
q_I=\sqrt{-g}G_{IJ}F^{J\,rt}.
\label{eq:consq}
\end{equation}

Substituting (\ref{eq:rttval}) and (\ref{eq:consq}) into the bulk
action, (\ref{eq:bulktt}), we arrive at
\begin{eqnarray}
I_{\rm bulk}&=&-\fft1{8\pi G_d}\int d^{d-1}x\int_{r_+}^{r_0}dr
\fft{d}{dr}\left(\sqrt{-h}\Theta_t^t+\fft12A^I_tq_I\right)\nonumber\\
&=&-\fft{\beta\omega_{d-2}}{8\pi G_d}\left.\left(\sqrt{-h}\Theta_t^t
+\fft12A_t^I q_I\right)\right|_{r_+}^{r_0},
\end{eqnarray}
where $r_+$ is the location of the horizon.
We must add to this the Gibbons-Hawking term
\begin{equation}
I_{\rm GH}=\fft1{8\pi G_d}\int_{\dM_0} d^{d-1}x\sqrt{-h}\Theta
=\fft{\beta\omega_{d-2}}{8\pi G_d}\sqrt{-h}\Theta.
\end{equation}
The resulting action is thus given by
\begin{equation}
\beta\,\Omega_{\rm reg}\equiv
I_{\rm reg}=\fft{\beta\omega_{d-2}}{8\pi G_d}\left(
\sqrt{-h}(-\Theta^t_t+\Theta)+\left.\sqrt{-h}\Theta^t_t\right|_{r+}
-\fft12\Phi^Iq_I\right),
\label{eq:uromeg}
\end{equation}
where $\Phi^I=A_t^I(r_{0})-A_t^I(r_+)$.  It is now clear that the
first term, proportional to $(-\Theta_t^t+\Theta)$, may be related to
the ADM energy, the second term may be related to the product of temperature
with entropy, and the last term gives directly the product $\Phi^IQ_I$ up
to charge normalization
\begin{equation}
Q_I=\fft{\beta\omega_{d-2}}{16\pi G_d}q_I.
\end{equation}
More explicitly, the $tt$ component of the regulated boundary stress
tensor, (\ref{eq:bst}), is
\begin{equation}
T_{tt}=\fft1{8\pi G_d}(\Theta_{tt}-\Theta h_{tt}),
\end{equation}
so that the ADM energy is
\begin{equation}
E_{\rm reg}=\fft{\omega_{d-2}}{8\pi G_d}\sqrt{-h}(-\Theta_t^t+\Theta).
\end{equation}
In addition, the entropy and temperature are given by
\begin{eqnarray}
S&=&\left.\fft1{4G_d}A\right|_{r_+}=\fft{\omega_{d-2}}{8\pi G_d}
\left.\left(2\pi e^{(d-2)B}r^{d-2}\right)\right|_{r_+},\nonumber\\
T&=&\left.\fft1{4\pi}e^{A-B}\fft{df}{dr}\right|_{r_+}.
\end{eqnarray}
Hence
\begin{equation}
TS=\fft{\omega_{d-2}}{8\pi G_d}\left.\left(\fft12e^{A+(d-3)B}r^{d-2}
\fft{df}{dr}\right)\right|_{r_+}
=-\left.\fft{\omega_{d-2}}{8\pi G_d}\sqrt{-h}\Theta_t^t\right|_{r_+}
\end{equation}
Here we have used the expression
\begin{equation}
\sqrt{-h}\Theta_t^t=-e^{A+(d-3)B}\left(r^{d-2}f\fft{dA}{dr}+\fft12
r^{d-2}\fft{df}{dr}\right),
\end{equation}
which is valid for the metric (\ref{eq:ssmet}).  Note that the first
term in $\Theta_t^t$ vanishes at the horizon (at least for a regular
horizon).

Combining the above expressions, and substituting into (\ref{eq:uromeg}),
we finally obtain the expected relation
\begin{equation}
\Omega_{\rm reg}=E_{\rm reg}-TS-\Phi^IQ_I.
\end{equation}
Note that the unrenormalized quantities $\Omega_{\rm reg}$ and
$E_{\rm reg}$ both diverge as we remove the regulator, $r_0\rightarrow\infty$.
However, this divergence is cancelled
by a counterterm action (\ref{ctaction}) which contributes equally with
$\Omega_{\rm ct}=E_{\rm ct}$.  Hence, the thermodynamic relation
(\ref{eq:thermo}) always holds identically, with or without counterterm
insertion, at least for the counterterm structures that we are interested
in.

\section{Conclusions}

In general, the notion of mass or energy in a gravitational system can be
rather difficult to define in a precise and useful manner.  Nevertheless,
rigorous definitions of energy and conserved charges are essential in
the application of black hole thermodynamics.  Here we have highlighted
a holographic approach, based on the Hamilton-Jacobi formalism, to dealing
with black holes in asymptotically AdS spacetimes.  In this approach,
conserved quantities (including the
mass) may be extracted from the boundary stress tensor, so long as the
gravitational action itself is regulated in an appropriate manner.  We
demonstrate, in particular, that the Hamilton-Jacobi method generates
the appropriate boundary counterterms for removing all divergences of
the on-shell action pertaining to stationary $R$-charged AdS black holes
in four, five, six and seven-dimensional gauged supergravities.

Although the importance of the boundary stress tensor method has been
realized for some time, and the notion of holographic renormalization
has been well developed, less attention has been given to systems with
a non-trivial matter sector.  In this paper, we have focused on
gravitational systems with long-range scalars, and have shown that
they may be treated in a uniform manner, regardless of spacetime
dimension or specific matter content.  Of the actual black holes we
have investigated, we note that the non-trivial scalar counterterms
(namely the non-constant parts of the superpotential $W$) are divergent
in four, finite in five, and vanishing in higher dimensions.  And yet
they all have a common origin, namely the Hamiltonian constraint
(\ref{H0eqn}) arising from the Hamilton-Jacobi analysis of the counterterms, 

It would of course be natural to apply the Hamilton-Jacobi counterterm
prescription developed here to the study of thermodynamics of other
interesting systems with non-trivial matter fields.  For example,
masses of rotating supersymmetric AdS$_5$ black holes
\cite{Gutowski:2004ez,Gutowski:2004yv} was recently considered in
\cite{Liu:2004it}.  It may also be of interest to apply the
above methods in examining the properties of the five-dimensional
black ring solutions
\cite{Emparan:2001wn,Elvang:2003yy,Elvang:2004ds,Gauntlett:2004qy}.

Finally, note that the boundary stress tensor contains information
not just on the energy of the system but also on general conserved
quantities corresponding to additional Killing symmetries.  In
particular, angular momentum along with the thermodynamics of
rotating solutions has been explored in
\cite{Hawking:1998kw,Awad:1999xx,Awad:2000aj,Malcolm} (see also
\cite{Klemm:2000gh,Klemm:2000vn,Gibbons:2004uw,Cvetic:2004hs,Cvetic:2004ny}).
For stationary solutions, the analysis of the previous section indicates
that any suitably chosen regulator will preserve the thermodynamic relation
(\ref{eq:thermoi}).  However, the introduction of angular momentum
yields additional complications meriting further study \cite{Malcolm}.
The full resolution of the first law of black hole thermodynamics in
the AdS/CFT context with rotation will certainly be an important
accomplishment with widespread implications.

\bigskip

\section*{Acknowledgments}

This material is based upon work supported by the National Science
Foundation under grant PHY-0313416 and by the US Department of Energy
under grant DE-FG02-95ER40899.  JTL and WS wish to acknowledge the
hospitality of the Khuri lab at the Rockefeller University, where part
of this work was completed.



\begin{thebibliography}{99}

\bibitem{Hawking:1982dh}
S.~W.~Hawking and D.~N.~Page,
{\sl Thermodynamics Of Black Holes In Anti-De Sitter Space},
Commun.\ Math.\ Phys.\  {\bf 87}, 577 (1983).

\bibitem{Witten:1998zw}
E.~Witten,
{\sl Anti-de Sitter space, thermal phase transition, and confinement in gauge
theories},
Adv.\ Theor.\ Math.\ Phys.\  {\bf 2}, 505 (1998) [hep-th/9803131].

\bibitem{Cvetic:1999ne}
M.~Cvetic and S.~S.~Gubser,
{\sl Phases of $R$-charged black holes, spinning branes and strongly coupled
gauge theories},
JHEP {\bf 9904}, 024 (1999) [hep-th/9902195].

\bibitem{Brown:1992br}
J.~D.~Brown and J.~W.~York,
{\sl Quasilocal energy and conserved charges derived from the gravitational
action},
Phys.\ Rev.\ D {\bf 47}, 1407 (1993).

\bibitem{Balasubramanian:1999re}
V.~Balasubramanian and P.~Kraus,
{\sl A stress tensor for anti-de Sitter gravity},
Commun.\ Math.\ Phys.\  {\bf 208}, 413 (1999) [hep-th/9902121].

\bibitem{Emparan:1999pm}
R.~Emparan, C.~V.~Johnson and R.~C.~Myers,
{\sl Surface terms as counterterms in the AdS/CFT correspondence},
Phys.\ Rev.\ D {\bf 60}, 104001 (1999) [hep-th/9903238].

\bibitem{Henningson:1998gx}
M.~Henningson and K.~Skenderis,
{\sl The holographic Weyl anomaly},
JHEP {\bf 9807}, 023 (1998) [hep-th/9806087].

\bibitem{Kraus:1999di}
P.~Kraus, F.~Larsen and R.~Siebelink,
{\sl The gravitational action in asymptotically AdS and flat spacetimes},
Nucl.\ Phys.\ B {\bf 563}, 259 (1999) [hep-th/9906127].

\bibitem{Taylor-Robinson:2000xw}
M.~M.~Taylor-Robinson,
{\sl More on counterterms in the gravitational action and anomalies},
hep-th/0002125.


\bibitem{deHaro:2000xn}
S.~de Haro, S.~N.~Solodukhin and K.~Skenderis,
{\sl Holographic reconstruction of spacetime and renormalization in the
AdS/CFT correspondence},
Commun.\ Math.\ Phys.\  {\bf 217}, 595 (2001) [hep-th/0002230].

\bibitem{Skenderis:2000in}
K.~Skenderis,
{\sl Asymptotically anti-de Sitter spacetimes and their stress energy tensor},
Int.\ J.\ Mod.\ Phys.\ A {\bf 16}, 740 (2001) [hep-th/0010138].

\bibitem{Bianchi:2001de}
M.~Bianchi, D.~Z.~Freedman and K.~Skenderis,
{\sl How to go with an RG flow},
JHEP {\bf 0108}, 041 (2001) [hep-th/0105276].

\bibitem{Bianchi:2001kw}
M.~Bianchi, D.~Z.~Freedman and K.~Skenderis,
{\sl Holographic renormalization},
Nucl.\ Phys.\ B {\bf 631}, 159 (2002) [hep-th/0112119].

\bibitem{Skenderis:2002wp}
K.~Skenderis,
{\sl Lecture notes on holographic renormalization},
Class.\ Quant.\ Grav.\  {\bf 19}, 5849 (2002) [hep-th/0209067].


\bibitem{deBoer:1999xf}
J.~de Boer, E.~Verlinde and H.~Verlinde,
{\sl On the holographic renormalization group},
JHEP {\bf 0008}, 003 (2000) [hep-th/9912012].

\bibitem{deBoer:2000cz}
J.~de Boer,
{\sl The holographic renormalization group},
Fortsch.\ Phys.\  {\bf 49}, 339 (2001) [hep-th/0101026].


\bibitem{Kalkkinen:2001vg}
J.~Kalkkinen, D.~Martelli and W.~Muck,
{\sl Holographic renormalisation and anomalies},
JHEP {\bf 0104}, 036 (2001) [hep-th/0103111].

\bibitem{Martelli:2002sp}
D.~Martelli and W.~Muck,
{\sl Holographic renormalization and Ward identities with the
Hamilton-Jacobi method},
Nucl.\ Phys.\ B {\bf 654}, 248 (2003) [hep-th/0205061].


\bibitem{Larsen:2003pf}
F.~Larsen and R.~McNees,
{\sl Inflation and de Sitter holography},
JHEP {\bf 0307}, 051 (2003) [hep-th/0307026].

\bibitem{Larsen:2004kf}
F.~Larsen and R.~McNees,
{\sl Holography, diffeomorphisms, and scaling violations in the CMB},
hep-th/0402050.


\bibitem{Duff:1999gh}
M.~J.~Duff and J.~T.~Liu,
{\sl Anti-de Sitter black holes in gauged $N = 8$ supergravity},
Nucl.\ Phys.\ B {\bf 554}, 237 (1999) [hep-th/9901149].

\bibitem{Sabra:1999ux}
W.~A.~Sabra,
{\sl Anti-de Sitter BPS black holes in N = 2 gauged supergravity},
Phys.\ Lett.\ B {\bf 458}, 36 (1999) [hep-th/9903143].

\bibitem{Cvetic:1999xp}
M.~Cvetic {\it et al.},
{\sl Embedding AdS black holes in ten and eleven dimensions},
Nucl.\ Phys.\ B {\bf 558}, 96 (1999) [hep-th/9903214].

\bibitem{Buchel:2003re}
A.~Buchel and L.~A.~Pando Zayas,
{\sl Hagedorn vs. Hawking-Page transition in string theory},
Phys.\ Rev.\ D {\bf 68}, 066012 (2003)
[hep-th/0305179].

\bibitem{Liu:2004it}
J.~T.~Liu and W.~A.~Sabra,
{\sl Mass in anti-de Sitter spaces},
hep-th/0405171.

\bibitem{Behrndt:1998ns}
K.~Behrndt, A.~H.~Chamseddine and W.~A.~Sabra,
{\sl BPS black holes in $N = 2$ five dimensional AdS supergravity},
Phys.\ Lett.\ B {\bf 442}, 97 (1998) [hep-th/9807187].

\bibitem{Behrndt:1998jd}
K.~Behrndt, M.~Cvetic and W.~A.~Sabra,
{\sl Non-extreme black holes of five dimensional $N = 2$ AdS supergravity},
Nucl.\ Phys.\ B {\bf 553}, 317 (1999) [hep-th/9810227].

\bibitem{Romans:tw}
L.~J.~Romans,
{\sl The F(4) Gauged Supergravity In Six-Dimensions},
Nucl.\ Phys.\ B {\bf 269}, 691 (1986).

\bibitem{Liu:1999ai}
J.~T.~Liu and R.~Minasian,
{\sl Black holes and membranes in AdS$_7$},
Phys.\ Lett.\ B {\bf 457}, 39 (1999) [hep-th/9903269].

\bibitem{Gutowski:2004ez}
J.~B.~Gutowski and H.~S.~Reall,
{\sl Supersymmetric AdS$_5$ black holes},
JHEP {\bf 0402}, 006 (2004) [hep-th/0401042].

\bibitem{Gutowski:2004yv}
J.~B.~Gutowski and H.~S.~Reall,
{\sl General supersymmetric AdS$_5$ black holes},
JHEP {\bf 0404}, 048 (2004) [hep-th/0401129].

\bibitem{Emparan:2001wn}
R.~Emparan and H.~S.~Reall,
{\sl A rotating black ring in five dimensions},
Phys.\ Rev.\ Lett.\  {\bf 88}, 101101 (2002) [hep-th/0110260].

\bibitem{Elvang:2003yy}
H.~Elvang,
{\sl A charged rotating black ring},
Phys.\ Rev.\ D {\bf 68}, 124016 (2003) [hep-th/0305247].

\bibitem{Elvang:2004ds}
H.~Elvang, R.~Emparan, D.~Mateos and H.~S.~Reall,
{\sl Supersymmetric black rings and three-charge supertubes},
hep-th/0408120.

\bibitem{Gauntlett:2004qy}
J.~P.~Gauntlett and J.~B.~Gutowski,
{\sl General concentric black rings},
hep-th/0408122.

\bibitem{Hawking:1998kw}
S.~W.~Hawking, C.~J.~Hunter and M.~M.~Taylor-Robinson,
{\sl Rotation and the AdS/CFT correspondence},
Phys.\ Rev.\ D {\bf 59}, 064005 (1999) [hep-th/9811056].

\bibitem{Awad:1999xx}
A.~M.~Awad and C.~V.~Johnson,
{\sl Holographic stress tensors for Kerr-AdS black holes},
Phys.\ Rev.\ D {\bf 61}, 084025 (2000) [hep-th/9910040].

\bibitem{Awad:2000aj}
A.~M.~Awad and C.~V.~Johnson,
{\sl Higher dimensional Kerr-AdS black holes and the AdS/CFT correspondence},
Phys.\ Rev.\ D {\bf 63}, 124023 (2001) [hep-th/0008211].

\bibitem{Malcolm}
G.~W.~Gibbons, M.~J.~Perry and C.~N.~Pope, in preparation.

\bibitem{Klemm:2000gh}
D.~Klemm and W.~A.~Sabra,
{\sl General (anti-)de Sitter black holes in five dimensions},
JHEP {\bf 0102}, 031 (2001) [hep-th/0011016].

\bibitem{Klemm:2000vn}
D.~Klemm and W.~A.~Sabra,
{\sl Charged rotating black holes in $5d$ Einstein-Maxwell-(A)dS gravity},
Phys.\ Lett.\ B {\bf 503}, 147 (2001) [hep-th/0010200].

\bibitem{Gibbons:2004uw}
G.~W.~Gibbons, H.~Lu, D.~N.~Page and C.~N.~Pope,
{\sl The general Kerr-de Sitter metrics in all dimensions},
hep-th/0404008.

\bibitem{Cvetic:2004hs}
M.~Cvetic, H.~Lu and C.~N.~Pope,
{\sl Charged Kerr-de Sitter black holes in five dimensions},
hep-th/0406196.

\bibitem{Cvetic:2004ny}
M.~Cvetic, H.~Lu and C.~N.~Pope,
{\sl Charged rotating black holes in five dimensional $U(1)^3$ gauged $N=2$
supergravity},
hep-th/0407058.

\end{thebibliography}
\end{document}